\documentclass{article}[11pt]

\usepackage{comment}
\usepackage{url}

\usepackage{a4wide}

\usepackage{mediabb}
\usepackage{graphics}

\usepackage{epsf}
\usepackage{color}
\usepackage{amsmath}
\usepackage{amssymb}
\usepackage{latexsym}
\usepackage{slashed}
\usepackage{float}
\usepackage{multirow}

\DeclareMathOperator{\BR}{BR}

\newcommand{\oldcolor}[1]{{\color{black}#1}}
\newcommand{\cred}[1]{{\color{black}#1}}

\newcommand{\fbinv}{\,\text{fb}^{-1}}
\newcommand{\GeV}{\,\text{GeV}}
\newcommand{\shat}{\hat{s}}

\newcommand{\vep}{\varepsilon}

\newcommand{\al}[1]{\begin{align}#1\end{align}}

\newcommand{\bp}{\begin{pmatrix}}
\newcommand{\ep}{\end{pmatrix}}
\newcommand{\bb}{\begin{bmatrix}}
\newcommand{\eb}{\end{bmatrix}}
\newcommand{\nn}{\nonumber\\}
\newcommand{\paren}[1]{\left(#1\right)}
\newcommand{\sqbr}[1]{\left[#1\right]}
\newcommand{\ab}[1]{\left|#1\right|}
\newcommand{\fn}[1]{\!\left(#1\right)}
\newcommand{\br}[1]{\left\{#1\right\}}

\newcommand{\te}{\text}

\newcommand{\Slash}[1]{{\ooalign{\hfil#1\hfil\crcr\raise.167ex\hbox{/}}}}
\newcommand{\beq}{\begin{equation}}
\newcommand{\eeq}{\end{equation}}

\newcommand{\MKK}{M_{\text{KK}}}

\begin{document}

\title{
A bound on Universal Extra Dimension Models\\
from \oldcolor{up to $2\fbinv$} of LHC Data at 7\,TeV
}
\author{
	\Large
	Kenji Nishiwaki,\thanks{
		E-mail: \tt nishiwaki@stu.kobe-u.ac.jp
		}
	{}
	Kin-ya Oda,\thanks{
		E-mail: \tt odakin@phys.sci.osaka-u.ac.jp
		}\smallskip\\
	\Large
	Naoya Okuda,\thanks{
		E-mail: \tt okuda@het.phys.sci.osaka-u.ac.jp
		}
	{} and
	Ryoutaro Watanabe\thanks{
		E-mail: \tt ryoutaro@het.phys.sci.osaka-u.ac.jp
		}
	\medskip\\
	$^*$\it Department of Physics, Kobe University, Kobe 657-8501, Japan\smallskip\\
	$^\dagger{}^\ddag{}^\S$\it Department of Physics,  
	Osaka University,  Osaka 560-0043, Japan
	\smallskip\\
	}

\maketitle
\begin{abstract}\noindent
The recent \oldcolor{up to $\sim 2\fbinv$} of data from the ATLAS and CMS experiments at the CERN Large Hadron Collider at 7\,TeV put an upper bound on the production cross section of a Higgs-like particle.
We translate the result\oldcolor{s of the $H\to WW \to l\nu l\nu$ and $H\to\gamma\gamma$ as well as} the combined analysis by the \oldcolor{ATLAS and} CMS into an allowed region for the Kaluza-Klein (KK) mass $M_{\text{KK}}$ and the Higgs mass $M_H$ for all the known Universal Extra Dimension (UED) models in five and six dimensions.
\oldcolor{Our bound is insensitive to the detailed KK mass splitting and mixing and hence complementary to all other known signatures.}
\end{abstract}
\vfill
\mbox{}\hfill KOBE-TH-11-07\\
\mbox{}\hfill OU-HET-722/2011
\newpage

\section{Introduction}
The ATLAS and CMS experiments at the CERN Large Hadron Collider (LHC) have presented their latest results for the $\oldcolor{\lesssim 2}\fbinv$ of data at the center of mass energy 7\,TeV
\oldcolor{at the XXV International Symposium on Lepton Photon Interactions at High Energies (Lepton Photon 11), Mumbai, India, 22--27 August 2011.}
One of the most remarkable among them is the bound on the Higgs mass in the Standard Model (SM).
A combined analysis of the ATLAS experiment excludes the existence of the SM Higgs 
in mass ranges \oldcolor{$146\GeV < M_H < 232\GeV$, $256\GeV < M_H < 282\GeV$, and $296\GeV<M_H<466\GeV$}
within the 95\% Confidence Level (CL) based on 1.0--$2.3\fbinv$ data~\oldcolor{\cite{ATLAS_Lepton_Photon}}
and 
that of the CMS experiment excludes
\oldcolor{$145\GeV < M_H < 216\GeV$\oldcolor{,} $226\GeV  < M_H < 288\GeV$, and $310\GeV<M_H<400\GeV$}
within the 95\%\,\text{CL} based on \oldcolor{1.1--$1.7\fbinv$} data~\oldcolor{\cite{CMS_Lepton_Photon}}.
Further the 
production cross section of a Higgs-like particle, a particle that decays the same way as the SM Higgs, is severely constrained by these data in the still-allowed regions\oldcolor{, namely light $115\GeV<M_H<145\GeV$, middle $288\GeV<M_H<296\GeV$, and heavy $M_H>466\GeV$ windows.}

In this Letter, we translate the above constraint on the production cross section into that on the Kaluza-Klein (KK) scale of various 5-Dimensional~(5D) and 6D Universal Extra Dimension (UED) models, namely the minimal UED (mUED)~\cite{Appelquist:2000nn} and the Dirichlet Higgs (DH)~\cite{Haba:2009pb,Nishiwaki:2010te} models in 5D and the 6D UED models on $T^2/Z_2$~\cite{Appelquist:2000nn}, $T^2/Z_4$~\cite{Dobrescu:2004zi,Burdman:2005sr}, $T^2/(Z_2 \times Z'_2)$~\cite{Mohapatra:2002ug}, ${RP}^2$~\cite{Cacciapaglia:2009pa}, $S^2/Z_2$~\cite{Maru:2009wu}, Projective Sphere (PS)~\cite{Dohi:2010vc}, and $S^2$~\cite{Nishiwaki:2011gm}.\footnote{
In~\cite{Dohi:2010vc} the terminology ``real projective plane'' is employed for a sphere with its antipodal points being identified. In order to distinguish \cite{Dohi:2010vc} from \cite{Cacciapaglia:2009pa}, we call the former the Projective Sphere (PS). \oldcolor{We note that the PS and $S^2$ UED models have no orbifold fixed point and hence no localized interaction on it.}
}
Concretely, we bound the UED parameter space of $\MKK$ \oldcolor{(first-level KK mass)} and $M_{H}$ \oldcolor{(zero mode Higgs mass)}
based on the \oldcolor{leading ATLAS and CMS constraints on the total cross section and on that of each channel~\cite{ATLAS_Lepton_Photon,CMS_Lepton_Photon}}.

One of the biggest advantages of the UED models is the existence of a natural Dark Matter (DM) candidate, the Lightest KK Particle (LKP)~\cite{Servant:2002aq}.
The 6D UED models have further advantages of the requirement of the number of generations to be (zero modulo) three~\cite{Dobrescu:2001ae} and the assurance of the proton stability~\cite{Appelquist:2001mj}.

There exist several bounds on the 5D mUED model, within which the brane-localized interactions are assumed to be vanishing at the 5D Ultra-Violet (UV) cutoff scale $\oldcolor{\Lambda_{5D}}$.
The latest analysis on DM relic abundance including the effects from second KK resonances gives the preferred KK scale at around $M_{\text{KK}}\sim 1.3\,\text{TeV}$~\cite{Belanger:2010yx}.
\oldcolor{
	It is noted that the first KK charged Higgs becomes the LKP when $M_H\gtrsim 240$--300\,GeV, depending on the KK scale~\cite{Kakizaki:2006dz}.
	}
The electroweak precision data suggests \oldcolor{that the KK scale should be $M_\text{KK}\gtrsim 800\GeV$ ($300\GeV\lesssim M_\text{KK}\lesssim 400\GeV$) at the 95\% CL for $M_H = 115$ (700)\,GeV~\cite{Appelquist:2002wb,Gogoladze:2006br,Baak:2011ze}.} The observed branching ratio of $B_d \rightarrow X_s \gamma$ confines the KK scale as $\MKK > 600\,\text{GeV}$~\cite{Haisch:2007vb} at the $95\%\,\text{CL}$.
Recent study puts a constraint 
${\MKK > \cred{600}\,\text{GeV}}$ for $10 < \Lambda_{\oldcolor{5D}}/M_{\text{KK}} < 40$ at the $95\%\,\text{CL}$ \cred{\{\}} \cite{Murayama:2011hj}\oldcolor{,} from the ATLAS SUSY search result in multijet$+ E_{\text{T}}^{\text{miss}}$ with \cred{$1\,\text{fb}^{-1}$} data~\cite{Aad:2011ib}.\footnote{
Inclusion of the decay channel into KK Higgs, if allowed, might significantly affect the result. We thank K.\ Tobioka on this point.
}
\cred{We see that current LHC bound from jets plus missing $E_T$ is not severe even for the most constraint mUED. This is because we have typically smaller mass splitting between the LKP and other new particles than the one between the lightest supersymmetric particle and other sparticle in the minimal supersymmetric standard model.}

We note that all of these bounds are strongly dependent on the mass splitting \oldcolor{and mixing} within the first KK level and therefore on the boundary mass structure which is derived from the above-mentioned assumption that all of them are zero at the 5D UV cutoff scale.
The bound on the KK scale put in this Letter is complementary to them in the sense that this is depending only on the Higgs mass. That is, our bound is \oldcolor{insensitive to} the boundary masses if they are smaller than the KK scale, as is necessary to have a higher dimensional picture at all.



\section{
Procedure to obtain the bound
\label{section_gluonfusion}
}

The ATLAS and CMS groups have shown the results for the combined analyses for the ratio \oldcolor{$\sigma^{95\%}_{pp\to H}/\sigma^{\text{SM}}_{pp\to H}$} as a function of the Higgs mass $M_H$, where \oldcolor{$\sigma^{95\%}_{pp\to H}$} is an upper bound on the production cross section of a particle that decays the same as the SM Higgs, at the 95\% CL~\cite{ATLAS_Lepton_Photon,CMS_Lepton_Photon}. In our case the constrained production cross section is that of the UED Higgs.
\oldcolor{In UED models, a process can be affected by KK-loops when it is loop-induced in the SM.}
\oldcolor{In particular, the dominant Higgs production channel via the gluon fusion process can be greatly enhanced, see~\cite{Nishiwaki:2011gm} and references therein.}
\oldcolor{For middle and heavy Higgs mass regions, the constraint is mainly from the $H\to WW$ and $ZZ$ channels, which are dominated by the tree-level SM processes and therefore the result of the combined analysis can be applied directly.}

\oldcolor{
For the light Higgs mass region, the severest bound on $\sigma_{pp\to H}^\text{95\%}/\sigma_{pp\to H}^\text{SM}$ comes from $H\to WW\to l\nu l\nu$ or $H\to\gamma\gamma$. 
The latter is loop-induced in the SM and can be affected by the KK-loops. Further, the loop-induced decay into gluons is not negligible in this region in computation of the total decay width. Therefore, we cannot trust the combined analysis which assumes that the branching ratios are not changed from the SM. In the light mass range, we apply the CMS bounds on $\sigma_{pp\to H\to \gamma\gamma}^\text{95\%}/\sigma_{pp\to H\to \gamma\gamma}^\text{SM}$ and $\sigma_{pp\to H\to WW}^\text{95\%}/\sigma_{pp\to H\to WW}^\text{SM}$~\cite{CMS_Lepton_Photon}.
}

\oldcolor{For the Higgs production,} 
we focus on the 
the gluon fusion process via the (KK) top quark loops, which is 
the dominant Higgs production channel in the SM and the UED models~\cite{Nishiwaki:2011gm,Petriello:2002uu,Maru:2009cu,Nishiwaki:2011vi}.
The parton level cross section of each model $\hat\sigma^{\text{model}}_{gg\to H}$ is 
\al{
\hat\sigma^{\text{model}}_{gg\to H}(\hat{s})
	&=	{\pi^2\over8M_H}\,
			\Gamma_{H\to gg}^{   { \text{model}   } }(M_H)\,
			{\delta(\shat-M_H^2)},
			\label{total_cross_section}
}
where
	$\hat{s}$ is the parton level center-of-mass-energy-squared and 
	$\Gamma_{H\to gg}^{   { \text{model}   } }$ is the partial decay width into a pair of gluons in each model: 
\al{
\Gamma^{\text{model}}_{H\to gg}(M_H)
	&=	\oldcolor{K}{\alpha_s^2\over8\pi^3}{M_H^3\over v_\text{EW}^2}\,\ab{J_t^{\text{model}}\fn{\oldcolor{M_H^2}}}^2\oldcolor{,}
}
\oldcolor{where $\alpha_s$ is the QCD coupling strength and $v_\te{EW}$ is the Higgs vacuum expectation value $\simeq 246\,\te{GeV}$ and $K$ is the $K$-factor to take into account the higher order QCD corrections, whose NNLO value is $\simeq 1.9$ at the relevant energies, see e.g.\ Ref.~\cite{Djouadi:2005gi}. \oldcolor{When we consider a ratio such as $\sigma_{pp\to H}^\text{95\%}/\sigma_{pp\to H}^\text{SM}$ from the gluon fusion process, the overall K-factor does not influence the result. However it contributes to the decay branching ratios of the light Higgs boson non-negligibly.} }

For each model, the loop function $J_t^{\text{model}}$ describes the contributions of all the \oldcolor{zero and KK modes for the top quark} in the triangle loops:
\begin{align}
 J_t^\text{SM} (\hat s)
 	&=  I\fn{ m_t^2 \over \hat s }, \label{result_of_SM}
\end{align}
\begin{align}
 J_t^\text{mUED} (\hat s)
 	&=  \br{ I\fn{ m_t^2 \over \hat s } +2 \sum_{n=1}^\infty \left( {m_t \over m_{t(n)}} \right )^2 I\fn{ {m_{t(n)}^2 \over \hat s} } },\label{result_of_mUED}  \\
 J_t^\text{DH}(\hat s)
 	&= \sqrt{2} \vep_1  \sqbr{ \left| I\fn{ m_t^2 \over \hat s } +2\sum_{n=1}^\infty \left( {m_t \over m_{t(n)}} \right )^2 I\fn{ m_{t(n)}^2 \over \hat s } \right |^2 
 +\left |2 \sum_{n=1}^\infty \left ( {m_t \over m_{t(n)} } \right )^2 \tilde I\fn{m_{t(n)}^2 \over \hat s } \right |^2 }^{1/2}, \label{result_of_DH}
\end{align}
\vspace{-6mm}
\begin{align} 
	{J_t^{T^2/Z_2}(\hat s) = J_t^{{RP}^2}(\hat s)}
		&=  \br{ I\fn{ m_t^2 \over \hat s } +2 \sum_{{m+n \geq 1 \atop \text{or\ } m=-n \geq 1}}  \left( {m_t \over m_{t(m,n)}} \right )^2 I\fn{ m_{t(m,n)}^2 \over \hat s } },
	\label{result_of_T2Z2}\\  
 {J_t^{T^2/Z_4}(\hat s)}
 	&=  \br{ I\fn{ m_t^2 \over \hat s } +2 \sum_{m\geq1, n\geq0} \left( {m_t \over m_{t(m,n)}} \right )^2 I\fn{ m_{t({m,n})}^2 \over \hat s } },
 \label{result_of_T2Z4}\\ 
	{J_t^{T^2/(Z_2\times Z'_2)}(\hat s)}
		&=	 \br{ I\fn{ m_t^2 \over \hat s } +2 \sum_{m\geq0,n \geq 0, \atop (m,n) \not= (0,0)}  \left( {m_t \over m_{t(m,n)}} \right )^2 I\paren{ m_{t(m,n)}^2 \over \hat s } },
	\label{result_of_T2Z2Z2}
\end{align}
\begin{align}	
 J_t^{S^2/Z_2}(\hat s)
 	&=  \br{ I\fn{ m_t^2 \over \hat s } +2 \sum_{j\geq1}\left( {m_t \over m_{t(j)}} \right )^2 n^\oldcolor{S^2/Z_2}(j)\,I\paren{ m_{t(j)}^2 \over \hat s } },
 \label{result_of_S2Z2}\\
 J_t^\text{PS}(\hat s) = {J_t^{S^2}(\hat s)}
 	&=  \br{ I\fn{ m_t^2 \over \hat s } +2 \sum_{j\geq1}\left( {m_t \over m_{t(j)}} \right )^2 (2j+1)\,I\fn{ m_{t(j)}^2 \over \hat s } },
 \label{result_of_PS}
\end{align}
where 
$I$ and $\tilde{I}$ are given by
\al{
I(\lambda)
	&=	\oldcolor{-2\lambda+\lambda(1-4\lambda)}
		\int_0^1{dx\over x}\ln\sqbr{{x(x-1)\over\lambda}+1-i\epsilon},\\
\tilde{I}(\lambda)
	&=	\oldcolor{(+\lambda)}
		\int_0^1{dx\over x}\ln\sqbr{{x(x-1)\over\lambda}+1-i\epsilon},
}
\oldcolor{
the $n^\text{model}(j)$ counts the number of degeneracy:
\begin{align}
n^{S^2/Z_2}(j)
	&=	\begin{cases}j+1,  \\ j, \end{cases} &
 n_\text{even}^\text{PS}(j)&= \begin{cases} 2j+1, \\ 0, \end{cases} &
 n_\text{odd}^\text{PS}(j)&= \begin{cases} 0, & \quad\text{for }j=\text{\oldcolor{even}}, \\ 2j+1, & \quad\text{for }j=\text{\oldcolor{odd}},\end{cases} 
\end{align}
and we write the KK top and $W$ masses ($X=t,W$)} 
\begin{align}
 m_{\oldcolor X(n)}
 	&\equiv \sqrt{m_{\oldcolor X}^2 + {\frac{n^2}{R^2}}}
 	= 		\sqrt{m_{\oldcolor X}^2 + n^2{\MKK^2}}, \\
 m_{\oldcolor X(m,n)}
 	&\equiv \sqrt{m_{\oldcolor X}^2 + {\frac{m^2+n^2}{R^2}}}
 	=		\sqrt{m_{\oldcolor X}^2 + {\paren{m^2+n^2} {\MKK^2} } }, \\
 m_{\oldcolor X(j)}
 	&\equiv \sqrt{m_{\oldcolor X}^2 + {\frac{j(j+1)}{R^2}}}
 	= 		\sqrt{m_{\oldcolor X}^2 + {\frac{j(j+1) {\MKK^2}}  {2} } },
\end{align}
with ${\MKK}$ being the first KK mass: $\MKK=1/R$ for the $S^1/Z_2$ (mUED), an interval (DH), and $T^2$-based compactifications (namely $T^2/Z_2$, $T^2/(Z_2\times Z_2')$, $T^2/Z_4$ and $RP^2$) and being ${\MKK}=\sqrt{2}/R$ for the $S^2$-based ones (namely $S^2/Z_2$, PS {and $S^2$}).
The range of the KK summation reflects the structure of each extra dimensional background.\footnote{
The origin of the factor 2 in front of each KK summation is the fact {that} there are both left and right handed (namely, vector-like) KK modes for each chiral quark zero mode corresponding to a SM quark.}

The factor $\sqrt{2} \vep_1$ in Eq.~(\ref{result_of_DH}) is equal to $2\sqrt{2}/\pi \sim 0.90$.
Readers who want more explanations on the above expressions should consult Ref.~\cite{Nishiwaki:2011gm}.

\oldcolor{
As is mentioned above, we compute the decay rate into a photon pair, following~\cite{Nishiwaki:2011vi}. The result is
\begin{equation}\begin{split}
 &\Gamma^\text{model}_{H \to \gamma \gamma}(M_H)
 = {G_F \over 8\sqrt 2 \pi} M_H^3 \cdot {\alpha^2 \over \pi^2} \left | J_W^\text{model}\fn{M_H^2} +{4 \over 3} J_t^\text{model}\fn{M_H^2} \right |^2,
\end{split}\end{equation}
where
\begin{align}
 J_W^\text{SM} (M_H^2)
 	&=  L\fn{ {1\over2},3,3,6,0;{M_W^2\over M_H^2} ,{M_W^2\over M_H^2} }, \\
 J_W^\text{mUED} (M_H^2)
 	&=  J_W^\text{SM} (M_H^2)+ \sum_{n=1}^\infty L\fn{{1\over2},4,4,8,1;{M_W^2\over M_H^2},{M_{W(n)}^2\over M_H^2} },
\end{align}
\begin{align}
 J_W^{T^2/Z_4}(M_H^2)
 	&=  J_W^\text{SM} (M_H^2)+ \sum_{m\geq 1,n\geq 0} L\fn{ {1\over2},5,4,10,1;{M_W^2\over M_H^2},{M_{W(m,n)}^2\over M_H^2} },\label{T2Z4_W}\\ 
 J_W^{T^2/(Z_2\times Z'_2)}(M_H^2)
	&=  J_W^\text{SM} (M_H^2)+ \sum_{m\geq 0,n\geq 0\atop (m,n)\neq(0,0)} L\fn{ {1\over2},5,4,10,1;{M_W^2\over M_H^2},{M_{W(m,n)}^2\over M_H^2} },\\ 
 J_W^{T^2/Z_2}(M_H^2)
 	&=  J_W^\text{SM} (M_H^2)+ \sum_{{m+n \geq 1 \atop \text{or\ } m=-n \geq 1}} L\fn{ {1\over2},5,4,10,1;{M_W^2\over M_H^2},{M_{W(m,n)}^2\over M_H^2} },  \\
 J_W^{{RP}^2}(M_H^2)
	&=  J_W^\text{SM} (M_H^2)+ \sum_{(m,n)}^{A} L\fn{ {1\over2},4,4,8,1;{M_W^2\over M_H^2},{M_{W(m,n)}^2\over M_H^2} }
		+ \sum_{(m,n)}^{B} L\fn{ 0,1,0,2,0;{M_W^2\over M_H^2},{M_{W(m,n)}^2\over M_H^2} },
\end{align}
\begin{align}
 J_W^{S^2/Z_2}(M_H^2)
 	&=J_W^\text{SM} (M_H^2)+ \sum_{j\geq 1} n^{S^2/Z_2}(j)\,L\fn{ {1\over2},5,4,10,1;{M_W^2\over M_H^2},{M_{W(j)}^2\over M_H^2} },\\
 J_W^{S^2}(M_H^2)
 	&=J_W^\text{SM} (M_H^2)+\sum_{j\geq 1} (2j+1)\,L\fn{ {1\over2},5,4,10,1;{M_W^2\over M_H^2},{M_{W(j)}^2\over M_H^2} },\\
 J_W^\text{PS}(M_H^2)
  	&=	J_W^\text{SM} (M_H^2)+ \sum_{j\geq 1} \Bigg [ n_\text{even}^\text{PS}(j)\,L\fn{ {1\over2},4,4,8,1;{M_W^2\over M_H^2},{M_{W(j)}^2\over M_H^2} } \nn
	&\phantom{=	J_W^\text{SM} (M_H^2)+ \sum_{j\geq 1} \Bigg [}
		+n_\text{odd}^\text{PS}(j)\,L\fn{ 0,1,0,2,0;{M_W^2\over M_H^2},{M_{W(j)}^2\over M_H^2} } \Bigg ],\label{PS_W}
\end{align}
with
\begin{align}
L(a,b,c,d,e;\lambda_1,\lambda_2)
 &=	a+b\lambda_1- \left[\lambda_1 \fn{c-d\lambda_2} -e\lambda_2 \right ] \int_0^1{dx\over x}\ln\sqbr{{x(x-1)\over \lambda_2 }+1-i\epsilon}.
\end{align}
\oldcolor{The $A$-summation for $RP^2$ are over the region that satisfies both $m\geq1$ and $n\geq1$ as well as over the ranges $(m,n)=(0,2), (0,4), (0,6), \dots$ and $(m,n)=(2,0), (4,0), (6,0), \dots$. Similarly, the $B$-summation are over $m\geq1$ and $n\geq1$ as well as over $(m,n)=(0,1), (0,3), (0,5), \dots$ and $(m,n)=(1,0), (3,0), (5,0), \dots$.}
The Dirichlet Higgs model only allows the heavy mass region in which $H\to\gamma\gamma$ is irrelevant and hence we do not compute the process for it.
}

In six dimensional UED models, KK summation in Eqs.~\eqref{result_of_T2Z2}--\eqref{result_of_PS} \oldcolor{and \eqref{T2Z4_W}--\eqref{PS_W}} must be terminated by a UV cutoff, for which we take the \oldcolor{maximum and minimum} possible values consistent with the Naive Dimensional Analysis (NDA), shown in Table~\ref{UEDcutoffvalues}.
Let us briefly explain this treatment hereafter. For more details, see Ref.~\cite{Nishiwaki:2011gm}.
Since the electroweak symmetry is broken by the Higgs mechanism in the SM and UED models (except for the DH model), the gluon fusion process is described by a dimension-six operator at lowest \oldcolor{in 4D point of view after KK expansion}.
This means that the calculation is UV logarithmic-divergent (convergent) in six (five) dimensions.\footnote{
Of course 5D UED is non-renormalizable and we have to introduce a cutoff scale in theory. However, this does not appear in our analysis because we can calculate the gluon fusion process with no UV divergence in 5D UEDs.
}
Therefore we need to put an upper limit of the summation over KK indices in 6D. We do it by adopting the NDA.
In both the $T^2$ and $S^2$-based geometries, the most stringent bound turns out to be the one from the perturbativity of the $U(1)_Y$ gauge interaction, which results in the following allowed regions of KK indices~\cite{Nishiwaki:2011gm}:
\al{
m^2 + n^2 &\lesssim \oldcolor{30},  &
&\text{for $T^2$-case} \ \paren{m_{(m,n)}^2 = \frac{m^2 + n^2}{R^2}},
\label{T2caseallowedindices} \\
j &\lesssim \oldcolor{9}, &
&\text{for $S^2$-case} \ \paren{m_{(j,m)}^2 = \frac{j(j+1)}{R^2}},
\label{S2caseallowedindices}
}
where the index $m$ for the $S^2$-case 
discriminates the degenerate states of each $j$-th level.
The cutoff scale of 6D UED theory $\Lambda_{6D}$ must be lower than that in Eq.~(\ref{T2caseallowedindices}) or (\ref{S2caseallowedindices}). In Table~\ref{UEDcutoffvalues}, we list the values that we take.

Based on the knowledge sketched above, we can evaluate the total cross section of the Higgs production of the UED models $\sigma_{pp\to H}^{\text{model}}$ and the ratio to that of the SM $\sigma_{pp\to H}^{\text{model}}/\sigma^{\text{SM}}_{pp\to H}$ to be compared to the experimental result. 

\section{Results
\label{section_results}}

\begin{table}
\begin{center}
\oldcolor{
\begin{tabular}{|c||cc|cc|}
\hline
& \multicolumn{2}{|c|}{$T^2$-based}
& \multicolumn{2}{|c|}{$S^2$-based}\\
& max & min & max & min \\
\hline
KK index & $m^2+n^2 < 28$ & $m^2+n^2 \leq 10$ & $j(j+1) \leq 90$ & $j(j+1) \leq 30$\\
UV cutoff & 
	$\Lambda_{6D} \sim 5\MKK$ & 
	$\Lambda_{6D} \sim 3\MKK$ & 
	$\Lambda_{6D} \sim 7\MKK$ & 
	$\Lambda_{6D} \sim 4\MKK$ \\
\hline
\end{tabular}
}
\caption{{\oldcolor{Our} choices of \oldcolor{maximum} and minimum upper bounds for KK indices and for the corresponding UV cutoff scale.}}
\label{UEDcutoffvalues}
\end{center}
\end{table}

\begin{figure}
\centering
\includegraphics[width=\textwidth, bb=0 0 940 320 , clip]{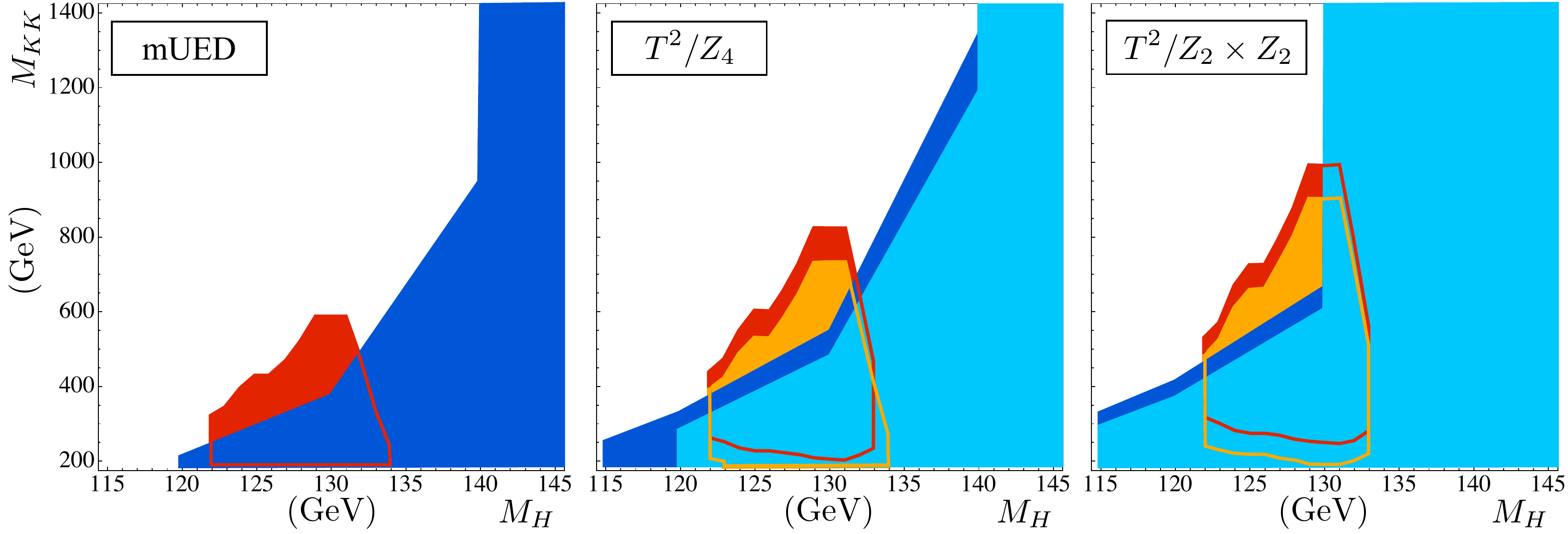}\\
\includegraphics[width=\textwidth, bb=0 0 940 320 , clip]{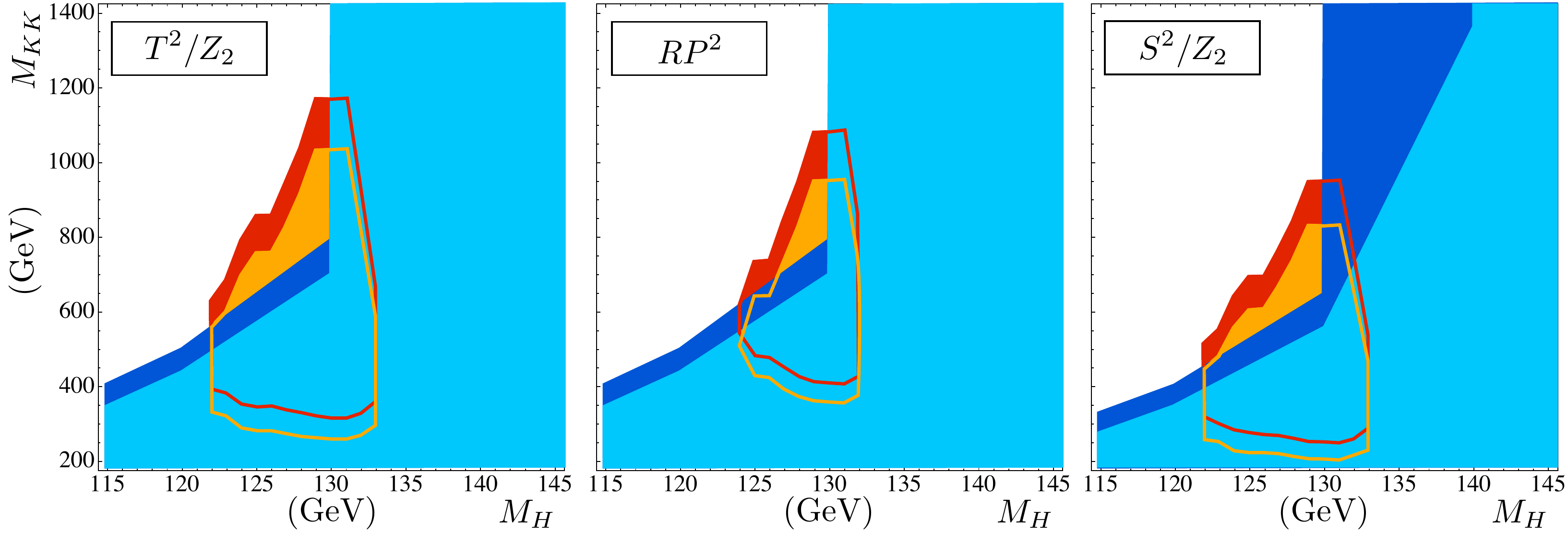}\\
\includegraphics[width=\textwidth, bb=0 0 938 320 , clip]{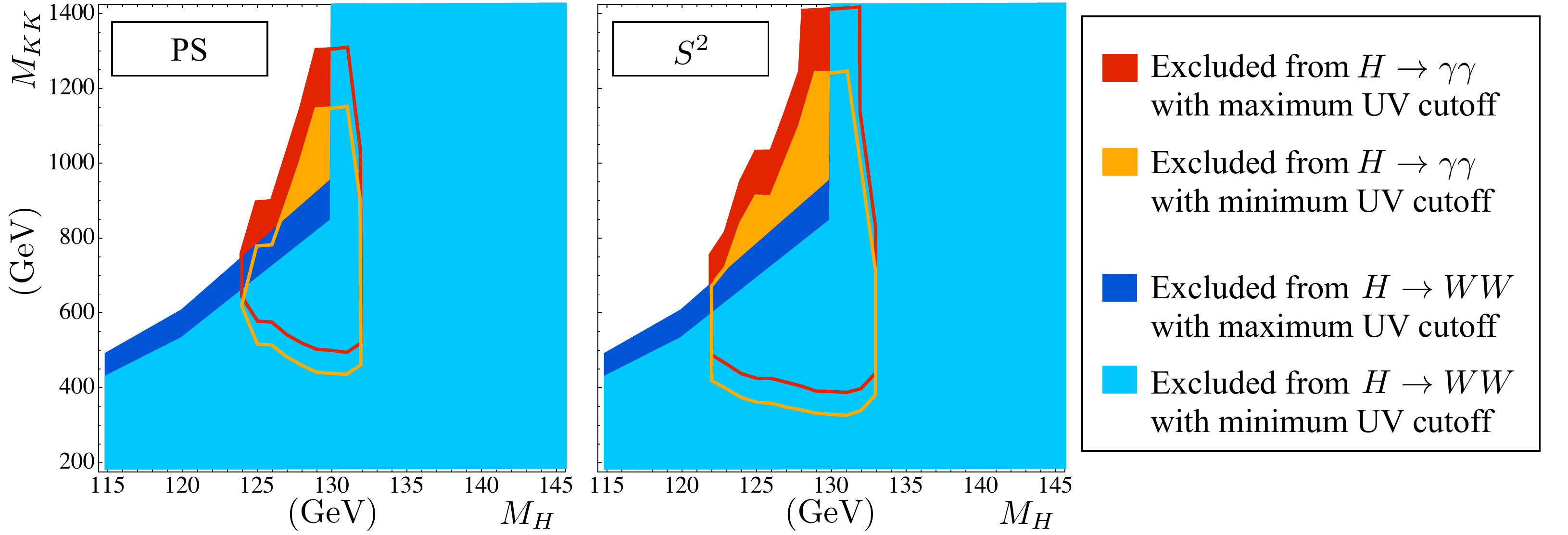}
\caption{
\oldcolor{Excluded} regions in $\MKK$-$M_H$ plane by the CMS constraints on \oldcolor{$\sigma^{95\%}_{pp\to H\to WW}/\sigma^{\text{SM}}_{pp\to H\to WW}$ and $\sigma^{95\%}_{pp\to H\to \gamma\gamma}/\sigma^{\text{SM}}_{pp\to H\to \gamma\gamma}$} for the \oldcolor{light} Higgs.
\oldcolor{
The bound from $WW$ channel (cyan and blue respectively for minimum and maximum UV cutoffs, former of which is superimposed on the latter) is superimposed on that from $\gamma\gamma$ channel (orange and red, the same as above) leaving its outline. The UV cutoff scales of our choice are summarized in Table~\ref{UEDcutoffvalues}.
}
}
\label{graph1}
\end{figure}

First, we show our results for the light region: $115\,{\text{GeV}}<M_H<145\,\text{GeV}$. 
We apply the CMS bounds on $H \to \gamma \gamma $ and $H \to WW$ channels that are dominant in this range.
In Fig.~\ref{graph1}, we list the contour plots for the \oldcolor{excluded} region in the $\MKK$-$M_H$ plane for various UED models \cred{in 5 and 6 dimensions}.
Plots for the maximum and minimum choices of the UV cutoff scale are presented for the 6D UED models.
In general, UED models enhance Higgs production via gluon fusion and reduce the Higgs decay into a pair of photons.
Therefore, $\sigma^\text{UED}_{pp\to H \to \gamma\gamma}$ receives nontrivial contributions from such effects.
\cred{Typically, the enhancement of Higgs production cross section overcomes the suppression of the di-photon branching ratio in the $H\to\gamma\gamma$ excluded range (with orange and red colors), whereas the region for smaller $M_\text{KK}$ is not excluded because of the suppression of the di-phton branching ratio. (For example, in the case of $S^2$ UED, the di-photon cross section is suppressed for $M_\text{KK}\lesssim 400\GeV$.) We find that all the suppressed region is already excluded by $WW$ channel.}
It is natural that the lower the cutoff scale becomes, the more the allowed parameter region is enlarged since smaller numbers of KK tops contribute to the process.
In 6D, we have more light KK top quarks running in the loop, and get stronger constraints than in 5D. 
The $\BR(H\to WW)$ is also affected by the enhancement of the total Higgs decay rate due to the increase of $H\to gg$.



\oldcolor{Second,} we move on to the \oldcolor{middle} region: \oldcolor{$288\GeV<M_H<296\GeV$.
The Standard Model is still allowed in this range whereas we find that all the UED models below $M_{KK} =1.4$ TeV is excluded.}



\begin{figure}
\centering
\includegraphics[width=\textwidth, bb=0 0 939 320 , clip]{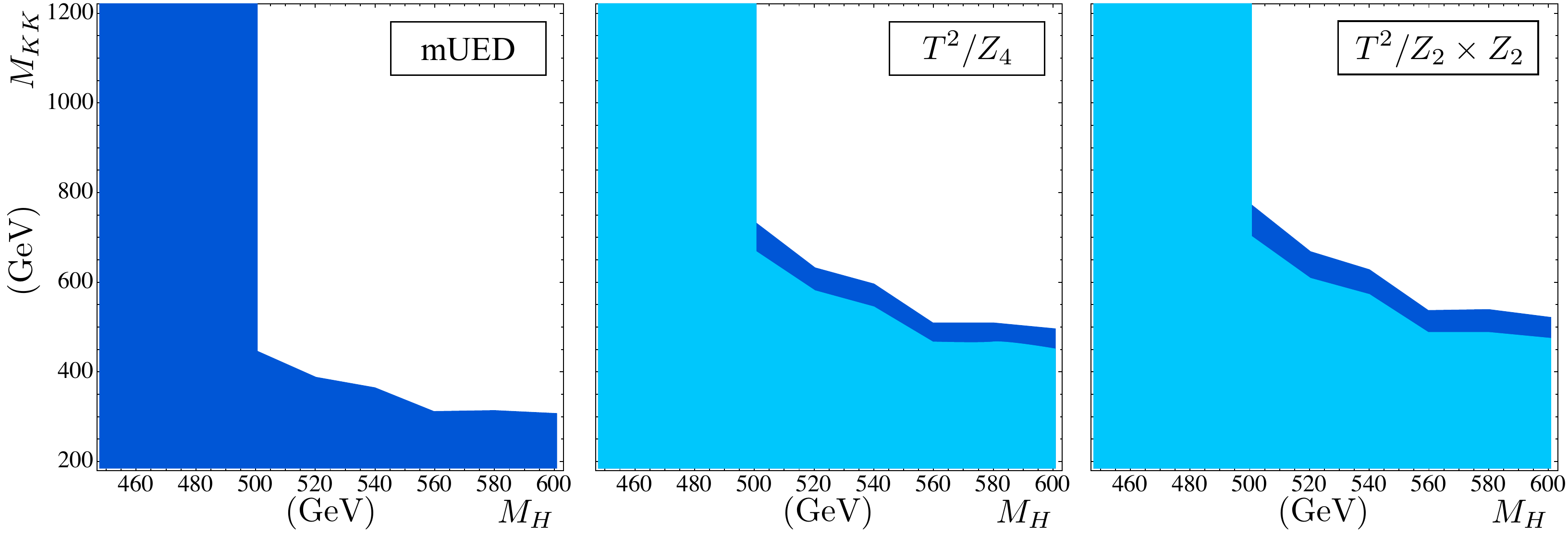}\\
\includegraphics[width=\textwidth, bb=0 0 939 320 , clip]{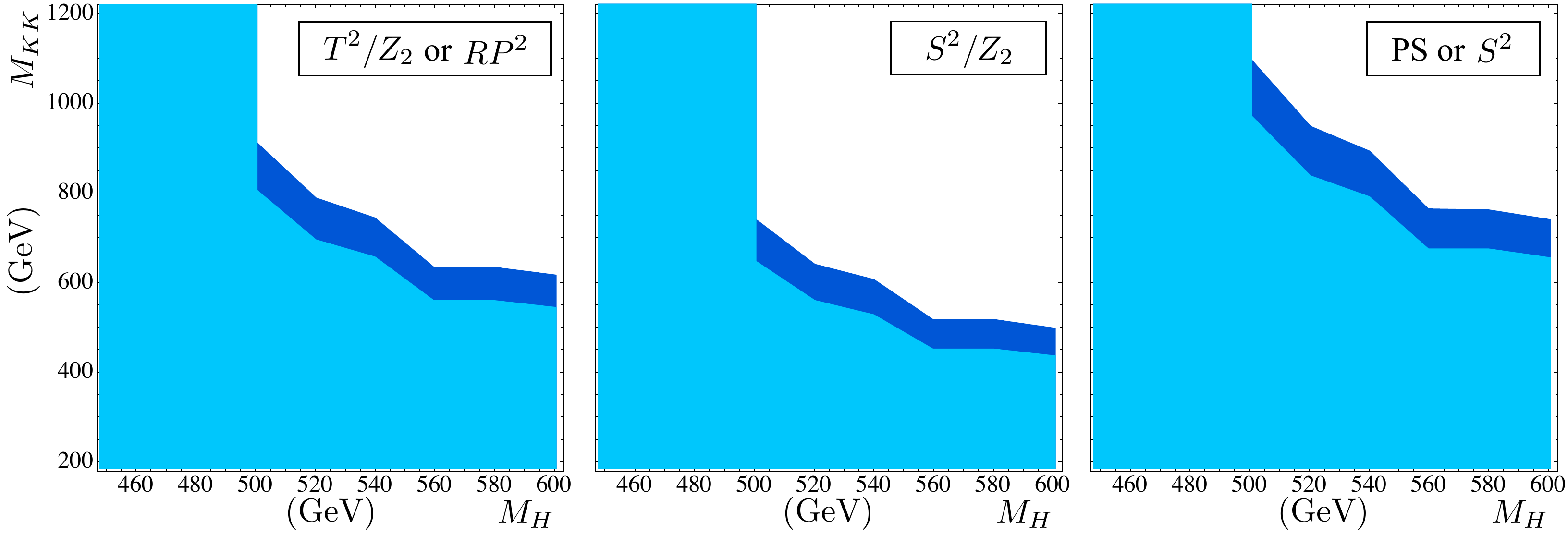}
\caption{
\oldcolor{
Constraints from the combined analysis from each ATLAS and CMS experiment in the heavy mass region, drawn the same as Fig.~\ref{graph1} with maximum UV cutoff (blue) being superimposed by the minimum UV cutoff (cyan).}
}
\label{graph2}
\end{figure}

Finally, let us discuss the \oldcolor{heavy} region: \oldcolor{$M_H > 446\GeV$.}
\oldcolor{We choose severer bound on $\sigma^\text{95\%}_{pp\to H}/\sigma^\text{SM}_{pp\to H}$ between ATLAS and CMS data for each $M_H$. That is, we use ATLAS and CMS bounds for $M_H<500\GeV$ and $M_H\geq 500\GeV$, respectively.}
In Fig.~\ref{graph2} we plot our results.
We note that in all the allowed region, we get $M_H<2\MKK$ and hence the Higgs does not decay into a pair of KK particles.

Now let us comment on the DH model.
In this model, the bound is put only on $\MKK$ ($= M_H$) and
the theoretical value of the ratio $\sigma_{pp\to H}^{\text{DH}}/\sigma^{\text{SM}}_{pp\to H}$ decreases when one increases $\MKK$, while the experimental upper bound $\sigma_{95\%}/\sigma_\text{SM}$ is an increasing function of $M_H$ in the high-mass region.
The cross-over occurs at $M_{\text{KK}} = 480\,\text{GeV}$ which gives $\sigma_{pp\to H}^{\text{DH}}/\sigma^{\text{SM}}_{pp\to H}\simeq 1.2$ and we  conclude that the allowed parameter region of $\MKK$ is:
\al{
M_{\text{KK}} > 480\,\text{GeV} \quad (\text{95\% CL in Dirichlet Higgs model}).
}
\noindent
In the $M_{\text{KK}}$ region of $110\,{\text{GeV}} < M_{\text{KK}} < 149\,\text{GeV}$ and $206\,{\text{GeV}} < M_{\text{KK}} < 300\,\text{GeV}$, the value of $\sigma_{pp\to H}^{\text{DH}}/\sigma^{\text{SM}}_{pp\to H}$ grows 
significantly and thereby these regions are rejected by the CMS result at the 95\% CL.
Noting that the indirect electroweak constraint gives $430\GeV < M_\text{KK} < 500\GeV$ at the 90\% CL~\cite{Haba:2009pb},
the allowed region of $\MKK$ roughly lies between $480\,\text{GeV} \lesssim M_{\text{KK}} \lesssim 500\,\text{GeV}$.

\section{Summary and Discussions
\label{section_summary}}
In this Letter we have constrained the UED models in 5D and 6D by use of the latest \oldcolor{ATLAS and} CMS bounds on the \oldcolor{Higgs} production cross section. 
The bound on 6D UED is severer than that on 5D UED because 6D KK mass spectrum is denser than that in 5D and therefore the KK top modes contribute to the gluon fusion process larger.
The KK (Higgs) mass of the Dirichlet Higgs model is pinned down at around 500\,GeV.

\oldcolor{
In the light mass range $115\GeV<M_H<145\GeV$, one of the dominant constrains on the Higgs production cross section is coming from the $H\to\gamma\gamma$ decay, which is reduced when KK scale is not large, due to the interference between the SM gauge boson and KK top loops, with some corrections from SM top and KK gauge bosons. We have taken into account the CMS constraints from $H\to\gamma\gamma$ and $H\to WW\to l\nu l\nu$ instead of the combined analysis. We find in this light region that the Higgs mass above 140\,GeV is already excluded in the 5D and 6D $T^2/Z_4$ UED models, while the mass above 130\,GeV is ruled out in other 6D UED models, both within a reasonably small KK scale $<1.4\,\text{TeV}$.
}


Our analyses on the 6D UED models cannot evade ambiguities from the NDA, but the plots in Figs.~\ref{graph1} and \ref{graph2} imply that the dependence on the cutoff is rather mild $\lesssim 10\%$.
For a low cutoff scale, there can also be contributions from higher dimensional operators that must be taken into account.
We have ignored these possible contributions in our analysis.

In both 5D and 6D cases, the bound is \oldcolor{insensitive to} the detailed boundary mass structure. In this sense this \oldcolor{constraint} is complementary to other ones such as the relic abundance of the LKP and the $M_{T2}$ analysis of the \oldcolor{decay of the colored KK} into the LKP.

When the KK scale is not much heavier than the weak scale $\simeq246\GeV$, the UED models tend to prefer much heavier Higgs mass than in the SM in order to cancel the KK top loops in the $T$-parameter. (This contribution has the same origin as the gluon fusion process discussed in this Letter.) In this regard it would be important to put an experimental bound for the Higgs mass beyond 600\,GeV.

\cred{There is the triviality bound if the Higgs is heavy and the UED scale is light which are being studied by the authors, along with the vacuum stability bound for the case of light Higgs, and will be presented in a separate publication.}

\subsection*{Acknowledgment}
We are grateful to \oldcolor{Abdelhak Djouadi,} Koichi Hamaguchi and \oldcolor{Kohsaku} Tobioka for helpful communications. 
We also thank the Yukawa Institute for Theoretical Physics for hosting the workshop ``Field Theory and String Theory,'' YITP-W-11-05, during which this work was initiated.
K.O.\ is partially supported by Scientific Grant by Ministry of Education and Science (Japan), Nos.~\oldcolor{23104009}, 23740192 and 20244028.

\bibliographystyle{TitleAndArxiv}
\providecommand{\bysame}{\leavevmode\hbox to3em{\hrulefill}\thinspace}



\newpage

\end{document}